\documentclass[conference]{IEEEtran}
\IEEEoverridecommandlockouts
\usepackage{cite}
\usepackage{amsmath,amssymb,amsfonts}
\usepackage{graphicx}
\usepackage{textcomp}
\usepackage{xcolor}
\usepackage{url}
\usepackage{times}
\usepackage{algorithmic}
\usepackage{algorithm}
\usepackage{epsfig}
\usepackage{subfigure}
\usepackage{nccmath} 
\usepackage{url}
\usepackage{multirow}
\usepackage{wrapfig}
\usepackage{subfigure}
\usepackage{color}
\usepackage{footnote}
\usepackage{rotating}
\usepackage{pdflscape}
\usepackage{fixmath}
\usepackage{amsthm}
\usepackage{bm}

\newcommand \xor{\mathbin{\oplus}}

\newtheorem{definition}{Definition}

\def\BibTeX{{\rm B\kern-.05em{\sc i\kern-.025em b}\kern-.08em
    T\kern-.1667em\lower.7ex\hbox{E}\kern-.125emX}}

\begin{document}

\title{Generalized Centrality Aggregation and Exclusive Centrality}

\author{\IEEEauthorblockN{Mostafa Haghir Chehreghani}
\IEEEauthorblockA{\textit{Department of Computer Engineering} \\
\textit{Amirkabir University of Technology (Tehran Polytechnic)}\\
Tehran, Iran \\
mostafa.chehreghani@aut.ac.ir}
% \and
% \and
% \and
}

\maketitle

\begin{abstract}
There are several applications that benefit from a 
definition of centrality which is applicable to sets of vertices,
rather than individual vertices.
% However, they provide only two specific cases of
% many possiblities for extending a centrality notion from vertices to the sets.
However, existing definitions might not be able to help us in answering
several network analysis questions.
In this paper, we study generalizing aggregation of centralities of individual vertices, to the centrality of the set consisting of these vertices.
In particular, 
% we discuss that in several case,
% more the existing extensions of
% centrality notions, it is useful to have some other centrality
% extensions for the sets. 
% Then
we propose {\em exclusive betweenness centrality},
% a novel extension of betweenness centrality to sets,
defined as the number of shortest paths passing over exactly
one of the vertices in the set,
and
% We 
discuss how this can be useful in determining the proper
center of a network.
% measure fullfits the conditions required for the centrality of
% a set, e.g., it says whether a set A is a good center of
% the network, or it should be extended to a larger set A 0 .
We mathematically formulate the relationship between
exclusive betweenness centrality and the existing notions of set
centrality, and use this relation to present an exact algorithm
for computing exclusive betweenness centrality.
Since
it is usually practically intractable to compute exact centrality
scores for large real-world networks, we also present approximate
algorithms for estimating exclusive betweenness centrality.
% In
% particlar, we present a general algorithm and discuss how it
% can be specialized to yield different source sampling, pair
% sampling and shortest path sampling algorithms.
In the end,
% by conducting experiments over several real-world networks,
% we epmirically evaluate our results. In our experiments, first
we evaluate the empirical efficiency of exclusive betweenness centrality computation over several real-world networks.
Moreover, we
empirically study the correlations between exclusive betweenness centrality and the existing set centrality notions.
\end{abstract}


\IEEEdisplaynotcompsoctitleabstractindextext
\IEEEpeerreviewmaketitle

\section{Introduction}

\textit{Centrality} is a structural attribute of vertices in a network,
used to determine the relative importance of a vertex or a set of vertices in the network.
For example, it can be used to determine how influential a person is within a social network, 
or how well-used a road/intersection is within a road network.
There are several measures of centrality in literature, including
\textit{degree centrality} defined as the number of links incident upon a vertex,
\textit{closeness centrality}, defined as
the inverse of the sum of the distances between the vertex and all other vertices, and
\textit{eigen-vector centrality} wherein
% It assigns relative scores to all nodes in the network based on the concept that 
connections to high-scoring vertices contribute
more to the centrality score of the vertex than equal connections to low-scoring vertices \cite{jrnl:Bonacich}.
Another well-known centrality notion is \textit{betweenness centrality},
% of a vertex is equal to 
defined as
the number (or the ratio) of shortest paths
from all vertices to all others that pass over the vertex.
It first was introduced by Linton Freemana as a measure for quantifying the control of a human
on the communication between other humans in a social network \cite{jrnl:Freeman}.

% \paragraph{Individual and group information centrality.}
There are several applications that benefit from a 
definition of centrality which is applicable to sets of vertices rather than individual vertices \cite{citeulike:392816}.
Therefore in the literature, a few extensions of 
centrality notions to sets are introduced.
When the centrality measure is betweenness centrality
(which is the main concern of this paper), two already studied ways of extending vertex betweenness centrality to sets of vertices are:
\begin{itemize}
\item \textit{group betweenness centrality} of a set,
which is defined as the number of shortest paths that pass over
\textbf{at least} one of the vertices in the set \cite{citeulike:392816,DBLP:conf/bigdataconf/ChehreghaniBA18}, and
\item \textit{co-betweenness centrality} of a set,
which is defined as the number of shortest paths that pass over
\textbf{all} the vertices in the set \cite{jrnl:Kolaczyk,DBLP:conf/wsdm/Chehreghani14}.
\end{itemize}

The authors of \cite{jrnl:Kolaczyk} showed that these two notions are related and developed a mathematical characterization of their relationship.
These two notions of {\em set centrality} provide a measure for the importance or influence of a set vertices in a network.
However, they provide only two specific cases of
many possibilities for extending a centrality notion from vertices to the sets.
Moreover, they might not be able to help us in answering
network analysis questions such as:
{\em when should a center consisting of an individual vertex $v$ be extended to a center consisting of $v$ and another vertex $u$?}
Or, more generally, how someone can decide {\em whether a set $A$ is a good enough center for a network $G$, or it should be extended to a larger set $A'$, $A \subset A'$,
in order to form a better center for $G$}?
As we will discuss in more details in Section \ref{sec:exclusive}, \textit{group betweenness centrality} and \textit{co-betweenness centrality}
do not provide proper answers to the above mentioned questions.

In this paper, we study generalizing aggregation of 
centrality of individual vertices to the sets.
In particular, we discuss that in several case,
more the existing extensions of centrality notions,
it is useful to have some other centrality extensions
for the sets.
Then we propose {\em exclusive betweenness centrality},
a novel extension of betweenness centrality to sets,
defined as the number of shortest paths passing over 
\textbf{exactly one} of the vertices in the set.
We discuss how this measure fulfills the conditions
required for the centrality of a set, e.g.,  
% This measure identifies the importance of a set vertice, 
% as well as 
it says whether a set $A$ is a good center of the network
or it should be extended to a larger set $A'$.
Then, We mathematically formulate the relationship between
exclusive betweenness centrality and the existing
notions of set centrality,
and use this relation to present
an exact algorithm for computing exclusive betweenness 
centrality.
Then, since it is usually practically intractable
to compute exact centrality scores in large real-world networks, we present approximate algorithms
for estimating exclusive betweenness centrality.
In particular, we present a general algorithm
and discuss how it can be specialized to yield
different {\em source sampling}, 
{\em pair sampling} and {\em shortest path sampling}
algorithms.
In the end, by conducting experiments over several real-world networks, we empirically evaluate our results.
In our experiments,
first we evaluate the running time of exact exclusive betweenness centrality computation over several
real-world networks.
Second, we empirically investigate the correlation
between exclusive betweenness centrality and group
betweenness centrality,
and the correlation between 
exclusive betweenness centrality and co-betweenness centrality.

The rest of this paper is organized as follows. 
In Section~\ref{preliminaries:sec}, preliminaries and definitions related to betweenness centrality are presented. 
In Section~\ref{sec:exclusive},
we motivate and introduce exclusive betweenness centrality
and discuss its usefulness.
In Section~\ref{sec:algorith}, we
discuss the mathematical relationship between
exclusive and co-betweenness centralities, and
present several exact and approximate algorithms
for computing/estimating exclusive betweenness centrality
of a given set (or all subsets of the vertices).
In Section~\ref{sec:experimentalresults},
we present our empirical results on 
exclusive betweenness centrality computation,
as well as on the correlations between
the set centrality notions.
In Section~\ref{section:relatedwork}, we have a brief overview on related work.
Finally, in Section~\ref{section:conclusion}, the paper is concluded.
% and a summary of the results presented in this paper is provided.

\section{Preliminaries}
\label{preliminaries:sec}

In this section, we present definitions and notations widely used in the paper.
We assume that the reader is familiar with basic concepts in graph theory.
Throughout the paper, $G$ refers to a graph (network).
For simplicity, we assume that $G$ is a connected and loop-free graph without multi-edges.
% In Sections \ref{sec:individual}-\ref{sec:sequence}, $\mathbf G$ points to an unweighted graph and in Section \ref{sec:weighted}, it points to a weighted graph.
$V(G)$ and $E(G)$ refer to the set of vertices and the set of edges of $
G$, respectively.
Furthermore, we use $n$ to point to $|V(G)|$,
and $m$ points to $|E(G)|$.
For an edge $e =(u,v) \in E(G)$,
$u$ and $v$ are the two end-points of $e$. 

% A graph $G'$ is a \textit{subgraph} of $G$ if $V(G') \subseteq V(G)$ and $E(G') \subseteq E(G)$.
% $G'$ is an \textit{induced subgraph} of $G$, if $V(G') \subseteq V(G)$ and $E(G')$ contains all edges of $E(G)$ which have both end-points in $V(G')$.

A \textit{shortest path} (also called a \textit{geodesic path}) between two vertices $u,v \in V(G)$ is a path
whose size is minimum, among all paths between $u$ and $v$.
For two vertices $u,v \in V(G)$, we use $d_{G}(u,v)$,
or $d(u,v)$ when $G$ is clear from the context,
to denote the size (the number of edges) of a shortest path
connecting $u$ to $v$.
By definition, $d_{G} (u,v)=0$ and $d_{G} (u,v) = d_{G} (v,u)$.
% The diameter of $G$, denoted $diam(G)$, is defined as  ...

For $s,t \in V(G)$, $\sigma_{st}$ denotes the number of shortest paths between $s$ and $t$;
% By definition, $\sigma_{ss}=1$.
and $\sigma_{st}(v)$ denotes the number of shortest paths between $s$ and $t$ that also pass through $v$. 
We have:
\[\sigma_{s}(v)=\sum_{t \in V(G) \setminus \{s,v\}}\sigma_{st}(v).\]
\textit{Betweenness centrality} of a vertex $v$ is defined as:
\begin{equation}
% \mathcal B(v)= \sum_{s,t \in V(G) \setminus \{v\}} \frac{\sigma_{st}(v)}{\sigma_{st}}
\mathcal B(v)= \sum_{s,t \in V(G) \setminus \{v\}} \sigma_{st}(v).
\end{equation}

A notion which is widely used for counting the number of shortest paths in a graph is the directed acyclic graph (DAG) containing all shortest paths starting from a vertex $s$ (see e.g. \cite{jrnl:Brandes}). In this paper, we refer to it as the \textit{shortest-path-DAG}, or \textit{SPD} for short, rooted at $s$.
For every vertex $s$ in a graph $G$, the \textit{SPD} rooted at $s$ is unique, and it can be computed in $O(m)$ time for unweighted graphs and in $O(m+n\log n)$ time for weighted graphs with positive weights \cite{jrnl:Brandes}.
% In a SPD $D$ rooted at $s$, the \textit{depth} of $v \in V(D)$ , denoted $dep_D(v)$, is defined as $d_{D}(s,v)$.

% In \cite{jrnl:Brandes}, the authors introduced the notion of the
The \textit{dependency score} of a vertex $s \in V(G)$ on a vertex $x \in V(G) \setminus \{s\}$ is defined as:
\begin{equation}
% \delta_{s\bullet}(v)=\sum_{t \in V(G) \setminus \{v,s\}} \frac {\sigma_{st}(v)}{\sigma_{st}} 
\delta_{s\bullet}(v)=\sum_{t \in V(G) \setminus \{v,s\}} \sigma_{st}(v).
\end{equation}

We have \footnote{When defining betweenness centrality, 
similar to \cite{RePEc:eee:phsmap:v:363:y:2006:i:1:p:89-95,10.1109/TVCG.2006.122,DBLP:conf/wsdm/Chehreghani14},
in this paper we do not divide $\sigma_{st}(v)$ by $\sigma_{st}$.
Another common definition of betweenness centrality
wherein $\sigma_{st}(v)$ is divided by $\sigma_{st}$
is as follows
\cite{jrnl:Brandes,DBLP:journals/cj/Chehreghani14,DBLP:conf/pakdd/ChehreghaniBA18,DBLP:conf/bigdataconf/ChehreghaniBA18a,DBLP:conf/cikm/ChehreghaniBA19}:
$$\mathcal B(v)= \sum_{s,t \in V(G) \setminus \{v\}} \frac{\sigma_{st}(v)}{\sigma_{st}}.
$$
In a similar way, 
\textit{dependency score} of $s \in V(G)$ on a $x \in V(G) \setminus \{s\}$ is defined as follows:
$$\delta_{s\bullet}(v)=\sum_{t \in V(G) \setminus \{v,s\}} \frac {\sigma_{st}(v)}{\sigma_{st}}.$$
}:
\begin{equation}
\mathcal B(v)=\sum_{s \in V(\mathbf G) \setminus \{v\}} \delta_{s\bullet}(v) 
\end{equation}

% The authors of \cite{jrnl:Brandes} showed that dependency scores of a source vertex on different vertices in the network can be computed using a recursive relation,
% defined as the following:
% \begin{equation}
% \delta_{s\bullet}(v)=\sum_{w:v \in P_s(w)} \frac{\sigma_{sv}}{\sigma_{sw}}(1+\delta_{s\bullet}(w))
% \label{eq:recursive}
% \end{equation}
% where $P_s(w)$ is defined as:
% \[\{ u \in V(\mathbf G): \{u,w\} \in E(\mathbf G) \wedge d_{\mathbf G}(s,v)=d_{\mathbf G}(s,u)+1 \}\]
% 
% As mentioned in \cite{jrnl:Brandes}, given the SPD rooted at $s$,
% dependency scores of $s$ on all other vertices can be computed in $O(m)$ time.   

% The algorithm presented in \cite{jrnl:Brandes} computes the betweenness centrality of every vertex $v$ by solving one single-source shortest-paths problem, in the way that
% at the end of each iteration, the dependencies of $s$ on every vertex $v$ are added to the centrality score of $v$.
% This method gives an $O(nm)$ algorithm for computing the betweenness centrality of every single vertex. 

%%%%%%%%%%%%%%%%%%%%%%%%%%%%%%%%%%%%%%%%%%%%%%%%%%%%%%%%%%%%%%%%%%%%%%%%%%%%%%%%%%%%%%
 
\section{Exclusive centrality}
\label{sec:exclusive}

In the literature, there exist algorithms that extend
a centrality notion defined for a single vertex to a set of vertices \cite{citeulike:392816,jrnl:Kolaczyk,DBLP:conf/wsdm/Chehreghani14,DBLP:conf/bigdataconf/ChehreghaniBA18}. This is done for different centrality notions,
and the usefulness of such extensions has been studied in different applications.
We refer to such extensions, in a general form,
as the {\em set aggregation} of the centrality notion.
Hence, an aggregation function is used to aggregate 
centrality scores of individual vertices in the set 
and yield the centrality score of the whole set. 
However, in the literature only some specific aggregation functions have been introduced and used.
Foe example, in {\em group betweenness centrality},
the aggregation function is defined as counting the number
(or the ratio) of shortest paths that pass over at least one of the vertices in the set \cite{citeulike:392816,DBLP:conf/bigdataconf/ChehreghaniBA18}.
in {\em co-betweenness centrality},
the aggregation function is defined as counting the number (or the ratio) of shortest paths that pass over all the vertices in the set \cite{jrnl:Kolaczyk,DBLP:conf/wsdm/Chehreghani14}.
These two extensions are only two specific forms, out of many other possibilities, that are introduced and used in the literature.

In this paper, we go beyond these two specific cases,
formulate the general form of aggregation,
and motivate and discuss a new case of set aggregation.
Let $A\subset V(G)$ be the set for which we want to
define the centrality notion.
Given a centrality notion $\mathcal C$ and an aggregation function $Agg$ defined over the set of vertices in $A$,
the $\mathcal C$ centrality of $A$ is defined as follows:
\begin{equation}
\mathcal C_{Agg}(A) = Agg(\mathcal C(v): v\in A) 
\end{equation}

In \cite{citeulike:392816}, the authors suggested considering centrality of a set of vertices as
the number (or the ratio) of shortest paths that pass through at least one of the vertices in the set and introduce the \textit{group betweenness centrality}.
More formally\footnote{This definition of group betweenness centrality is consistent with our definition of
betweenness centrality. 
Similar to betweenness centrality,
another common definition of group betweenness centrality
is as follows:
\begin{equation*}
\mathcal{GB}(A)= \sum_{s,t \notin A: s\neq t}
\frac{\sigma^*_{st}(A)}{\sigma_{st}}.
% \sigma^*_{st}(A).
\end{equation*}
}:
\begin{equation}
\label{eq:groupbetweenness}
\mathcal{GB}(A)= \sum_{s,t \notin A: s\neq t}
% \frac{\sigma^*_{st}(A)}{\sigma_{st}},
\sigma^*_{st}(A),
\end{equation}
where $\sigma^*_{st}(A)$ refers to the number of shortest paths between $s$ and $t$ that pass through \textbf{at least one} of the vertices in $A$.
The other natural extension is \textit{co-betweenness centrality},
which is presented by Kolaczyk et.al. \cite{jrnl:Kolaczyk} as the following\footnote{Another common definition of
co-betweenness centrality
is as follows:
\begin{equation*}
\mathcal{CB}(A)= \sum_{s,t \notin A: s\neq t}
\frac{\hat{\sigma}_{st}(A)}{\sigma_{st}}.
% \hat{\sigma}_{st}(A),
\end{equation*}

}:
\begin{equation}
\label{eq:cobetweenness}
\mathcal{CB}(A)= \sum_{s,t \notin A: s\neq t}
% \frac{\sigma_{st}(A)}{\sigma_{st}},
\hat{\sigma}_{st}(A),
\end{equation}
where $\hat{\sigma}_{st}(A)$ is defined as
the number of shortest paths between $s$ and $t$ that pass through \textbf{all} the vertices in $A$,
% By $\sigma_{st}(A)$, $A \subseteq V(G)$, we denote
% the number of shortest paths between $s$ and $t$ that also pass through all vertices in $A$.
and
\[\hat{\sigma}_{s}(A)=\sum_{t \in V(G) \setminus \{s\} \setminus A}\hat{\sigma}_{st}(A).\]

In Equation \ref{eq:groupbetweenness},
for every two subsets $A$ and $A'$ of the vertices such that $A\subset A'$, we have: $\mathcal{GB}(A) \leq \mathcal{GB}(A')$,
i.e., by adding a new vertex to a set $A$, its group betweenness centrality increases or at least it does not change.
We may refer to this property of group betweenness centrality as
the {\em monotonicity} property.
This property makes it difficult to 
use group betweenness centrality for
answering questions such as the following:
% In this paper, instead of finding {prominent groups},
% we focus on the following question:
% \paragraph*{} 

{\em
Given a set $A \subset V(G)$ and a vertex $v \in V(G) \setminus A $,
between $A$ and $A \cup \{ v \}$ which one is a {\em better center} for $G$?
}

The reason is that on the one hand $A \cup \{ v \}$ has always a non-less group betweenness centrality than $A$.
On the other hand,
however it is always desirable to keep the center
of a network as small as possible.
So in this sense, sometimes we may prefer $A$ to $A \cup \{ v \}$
if adding the new vertex $v$ increases the centrality/importance of the set just a little (if not at all).
More than the applicability aspects, the computation cost may also
encourage us to choose small sets, as computing group betweenness
centrality of a set usually increases by increasing its size.

An attempt to find a good set of vertices as the {\em center}
of a network was finding a {\em prominent set} \cite{jrnl:PuzisAIComm}.
% This means when 'group betweenness centrality' is used, as has been discussed in ,
A {\em prominent set} is a set of minimum size,
such that every shortest path in the network passes through at least one of the vertices in the set.
However, this notion of \textit{prominent groups} has shortcomings.
First, the problem of finding a {\em prominent group} is a simple reduction
of the {\em minimal vertex cover} problem \cite{jrnl:PuzisAIComm} and hence, it is an NP-hard problem.
Second, the size of a \textit{prominent group} can be large,
as it tries to \textit{control} all the {\em flows}
(which are done through shortest paths) in the network. 
% We might be interested in small sets of vertices as centers, due to complexity as well as applicability reasons.
% The complexity of finding a \textit{prominent group} is exponential in terms of its size, even in the heuristic methods presented in  \cite{jrnl:PuzisAIComm}.

To answer the above-mentioned question and alleviate
the discussed challenges, we present a new set centrality aggregation function, that compares/ranks a set {\em against its subsets}.
In the proposed measure,
called {\em exclusive betweenness centrality} and denoted with 
$\mathcal {XB}$, the following observations are considered:
\begin{enumerate}
\item
Let $A$ be a subset of $V(G)$, $v$ be a vertex in $G$,
$S$ be the set of all shortest paths in $G$,
$S(v)$ be the set of shortest paths in $G$ that pass through $v$
and $S(A)$ be the set of shortest paths in $G$ that pass through at least one of members of $A$.
For two vertices $v_1, v_2 \in V(G) \setminus A$, 
if $|S(\{v_1\} \cup A)| > | S(\{v_2\} \cup A)|$,
the {\em exclusive betweenness centrality} of $ \{ v_1 \} \cup A$ should be greater than the {\em exclusive betweenness centrality} of $\{ v_2\} \cup A$.
\item 
If a {\em considerable number} of (or most of) those shortest paths of $G$ that pass over $v$ also pass over the members of $A$,
{\em exclusive betweenness centrality} of $A$ should be greater than {\em exclusive betweenness centrality} of $A \cup \{ v\}$.
The reason is that while computing centrality of $A \cup \{ v\}$ is more time consuming than computing centrality of  $A$,
$A \cup \{v\}$ does not control flows of the network (much) more than $A$.
This means finding larger sets as the centers of the network
must be done only when they considerably increase the control over the flows in the network (i.e., they have a considerably larger centrality score
than their proper subsets). 
\end{enumerate}

The first observation defines a property desirable when 
two different vertices $v_1$ and $v_2$ are added to the set.
The second observation defines a property desirable
when a new vertex is added to the set, compared to the case
wherein the new vertex is not added.
In the following,
first in Definition~\ref{def:xb}
we present the definition of exclusive betweenness centrality of a set.
% of vertices.
Then, we discuss that it satisfies the two above-mentioned properties.

% An immidiate definition for $\mathcal{XB}$ of a set of vertices which satisfies Equation \ref{eq:observation1} is pesented in

\begin{definition}
\label{def:xb}
Let $A=\{v_1 \hdots v_k \}$ be a set of vertices.
The {\em exclusive betweenness centrality} 
of $A$ is defined as follows:
\begin{equation}
\mathcal{XB} (\{v_1 \hdots v_k \}) =
|S(v_1) \xor \hdots \xor S(v_k) |
\label{def:exclsive1}
\end{equation} 
where $\xor$ of two sets $S_1$ and $S_2$ is their {\em exclusive or},
i.e.,
$S_1 \xor S_2 = (S_1 \setminus S_2) \cup (S_2 \setminus S_1) $,
and $S(v)$ is the set of shortest paths that pass through vertex $v$.
\end{definition}

\begin{figure}
\centering
\subfigure[A graph $G$]
{
\includegraphics[scale=0.5]{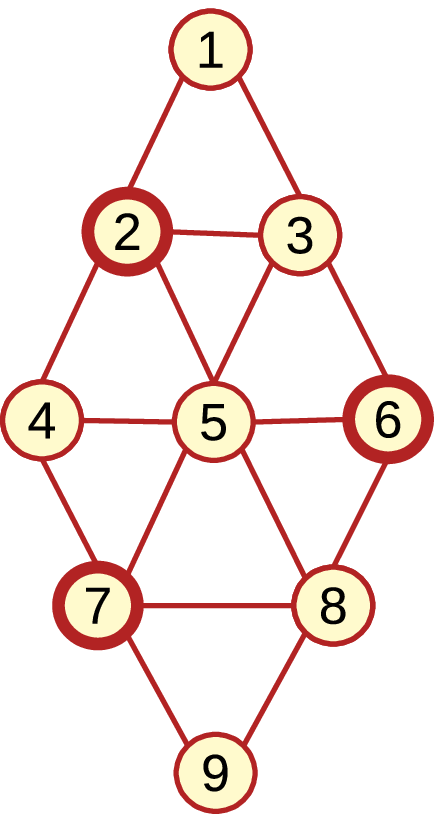}
\label{fig:example-a}
}
\subfigure[The shortest path DAG rooted at vertex $1$.]
{
\includegraphics[scale=0.5]{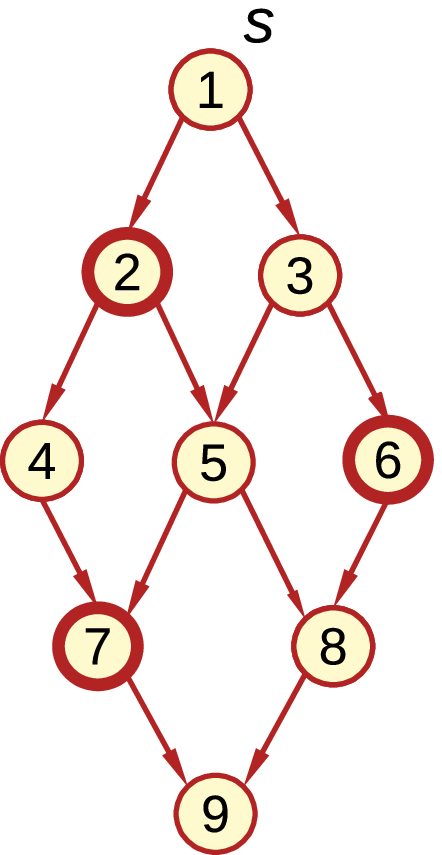}
\label{fig:example-b}
}
\caption
{
\label{fig:example}
An example of exclusive betweenness centrality.
Exclusive betweenness centrality of $A=\{2,6,7\}$ is $7$.
}
\end{figure}

As an example of exclusive betweenness centrality,
consider Figure~\ref{fig:example},
wherein Figure~\ref{fig:example-a} shows a graph $G$
and Figure~\ref{fig:example-b} shows the shortest path DAG
rooted at the vertex $s=1$.
Assume that the set $A$ consists of vertices $2,6,7$.
Those shortest paths that start from vertex $1$ and pass
over exactly one of the members of $A$
(hence, contribute to the exclusive betweenness centrality of $A$)
are as follows:
$1\rightarrow2\rightarrow4$,
$1\rightarrow2\rightarrow5$,
$1\rightarrow2\rightarrow5\rightarrow8$,
$1\rightarrow2\rightarrow5\rightarrow8\rightarrow9$,
$1\rightarrow3\rightarrow5\rightarrow7\rightarrow9$,
$1\rightarrow3\rightarrow6\rightarrow8$, and
$1\rightarrow3\rightarrow6\rightarrow8\rightarrow9$.
Therefore, exclusive betweenness centrality of $A$ is $7$.

The first observation mentioned above 
says that if the new vertex brings more {\em new} shortest paths controlled by the set, its union with the set must have
a larger centrality score. Therefore,
for any two vertices $v_1$ and $v_2$ and set $A$ of vertices,
we may express it as follows:
% can be expressed more formally,
% using the following property for $\mathcal {XB}$: 
% We can complete the first observation by defining the following property for $\mathcal {XB}$:
% The first observation, suggests the following property for $\mathcal {XB}$:

if
\begin{align*}
\left| \left(S(v_1) \setminus S(A) \right) \cup
\left( S(A) \setminus S(v_1) \right) \right| > \newline \\
\left| \left( S(v_2) \setminus S(A) \right) \cup
\left( S(A) \setminus S(v_2) \right) \right| 
\end{align*}

then
\begin{equation}
\mathcal{XB} (A \cup \{ v_1\} ) > \mathcal{XB} (A \cup \{ v_2\} )
\label{eq:observation1}
\end{equation}

It is easy to see that the definition of $\mathcal{XB}$
presented in Definition~\ref{def:xb} satisfies this property.

For the second observation, we need to define a threshold for
$|S(A \cup \{ v\}) \setminus S(A)|$
and as a result, for
$|S(v) \setminus S(A)|$.
We may define this threshold as the following:
\begin{equation}
\label{eq:threshold}
|S(v) \setminus S(A)| \geq |S(A \cap \{ v\})|. 
\end{equation}
% As the following analysis shows,
% This implies that
If the threshold of Equation~\ref{eq:threshold}
holds,
the number of shortest paths controlled by
$A \cup \{ v\}$ will be greater than or equal to the number of shortest paths controlled by $A$.
% According to the threshold of Equation~\ref{eq:threshold}
% We have:
% \begin{equation*}
% |S(A \cup \{ v\}) \setminus S(A)| \geq |S(A \cup \{ v\}) \cap S(A)|
% \end{equation*}
% which yields
% \begin{equation*}
% |S(v) \setminus S(A)| \geq |S(v) \cap S(A)|.
% \end{equation*}
More precisely, using some simple tricks from relational algebra, 
Equation~\ref{eq:threshold} yields:
\begin{equation}
|S(v) \setminus S(A) \cup S(A) \setminus S(v)  | \geq | S(A)|,
\label{eq:observation2}
\end{equation}
which yields that
\begin{equation}
\mathcal{XB}(A \cup \{v\}) \geq \mathcal{XB}(A).
\label{eq:observation3}
\end{equation}

This means if we use exclusive betweenness centrality,
as presented in Definition~\ref{def:exclsive1},
the larger set $A \cup \{v\}$ is preferred to
the smaller set $A$,
if the the threshold of Equation~\ref{eq:threshold}
holds, i.e., if the number of shortest paths controlled by $v$
but not by $A$ is greater than (or equal to) the number of shortest
paths controlled by both $A$ and $v$.

% Similar to Equation \ref{eq:observation1}, An immidiate definition for $\mathcal{XB}$ of a set of vertices which satisfies Equation \ref{eq:observation2} is Equation \ref{def:exclsive1}.
% the \textit{exclsive betweenness centrality} of a set $A \subseteq V(\mathbf G)$,
% defined as the number of shortest paths in the network that pass through exactly one vertex in $A$,
% satisfies the properties desired for the centrality of a set.

\section{Algorithmic aspects}
\label{sec:algorith}

In this section, we investigate
the relationship between {\em exclusive betweenness centrality}
and {\em co-betweenness centrality}, 
and exploit this relationship to present an exact algorithm for computing exclusive betweenness centrality.
We also discuss some approximation techniques
for estimating exclusive betweenness centrality.
% and finaly, we empirically analyze it.

\subsection{Relation to other centrality notions}

Kolaczyk et.al.~\cite{jrnl:Kolaczyk} showed that \textit{group betweenness centrality} of a set $A$ can be expressed
in terms of co-betweenness centrality, 
as follows\footnote{Note that in \cite{jrnl:Kolaczyk}, betweenness, group betweenness and co-betweenness centralities
are defined as the ratio of shortest paths that pass over a vertex/set. Since in this paper we define these centralities
as the number of shortest paths passing over a vertex/set,
we accordingly revise Equation~\ref{eq:cc}.
The original form of Equation~\ref{eq:cc} as presented in~\cite{jrnl:Kolaczyk} is as follows: 
\begin{equation*}
% \label{eq:cc}
\mathcal{CC}_A(\mathbf i_j) =\sum_{s,t \notin A: s \neq t } \frac{\sigma_{st}(\mathbf i_j)}{\sigma_{st}}.
\end{equation*}
}:
\begin{equation}
\mathcal{GB}(A)= \sum_{j=1}^{|A|} (-1)^{j-1} \sum_{\mathbf i_j \subseteq A} \mathcal {CC}_A( \mathbf i_j ),
\label{eq:relation1}
\end{equation}
where $\mathbf i_j$ is a subset of size $j$ of $A$ and $\mathcal {CC}_A(\mathbf i_j)$ is the $j$-th order co-betweenness of $\mathbf i_j$ with respect to $A$, defined as follows:
\begin{equation}
\label{eq:cc}
\mathcal{CC}_A(\mathbf i_j) =\sum_{s,t \notin A: s \neq t }
\hat{\sigma}_{st}(\mathbf i_j).
% \frac{\sigma_{st}(\mathbf i_j)}{\sigma_{st}}
\end{equation}

In a similar way, it can be shown
that \textit{exclusive betweenness centrality} can be re-written in terms of \textit{co-betweenness centrality}
of subsets of $A$ as follows:
\begin{equation}
\mathcal{XB} (A)= \sum_{j=1}^{|A|} j(-1)^{j-1} \sum_{\mathbf i_j \subseteq A} \mathcal{CC}_A( \mathbf i_j )
\label{eq:relation2}
\end{equation}
Note that for the special case of $|\mathbf i_j|=1$,
$\mathcal{CC}_A( \mathbf i_j )$ gives
betweenness centrality of $\mathbf i_j$,
with a small difference that source and target vertices of
shortest paths can not be in $A$.
We refer to this value as $\mathcal {B}_{A}(\mathbf i_j)$.
As a simple case, consider the situation wherein the set $A$ consists of two vertices $v_1$ and $v_2$.
Using Equation~\ref{eq:relation2}, 
\textit{exclusive betweenness centrality} of $\{v_1,v_2\}$
can be written as follows:
% is $|\mathcal S(v_1) \xor \mathcal S(v_1)|$.
% $\mathcal{XB}(\{v_1,v_2\})$ can be written as
\begin{align*}
\mathcal{XB}(\{v_1,v_2\}) = & \\ \nonumber
\mathcal {B}_{\{ v_1,v_2\}}(\{ v_1\}) + \mathcal {B}_{\{ v_1,v_2\}}(\{ v_2\}) - 2  \mathcal{CC}_{\{ v_1,v_2\}}( \{ v_1,v_2\} ) & .
\end{align*}
% $$\mathcal{XB}(\{v_1,v_2\})=$$
% $$\mathcal {BC}_{\{ v_1,v_2\}}(\{ v_1\}) + \mathcal {BC}_{\{ v_1,v_2\}}(\{ v_2\}) - 2 \mathcal C_{\{ v_1,v_2\}}( \{ v_1,v_2\} ).$$
% Note that the difference between 
% $\mathcal {B}_{\{ v_1,v_2\}}(\{ v_1\})$
% and 
% $\mathcal {B}(\{ v_1\})$
% is that unlike $\mathcal {B}(\{ v_1\})$,
% in $\mathcal {B}_{\{ v_1,v_2\}}(\{ v_1\})$
% the source and target vertices of the shortest paths
% cannot be $v_2$
% (in both $\mathcal {B}_{\{ v_1,v_2\}}(\{ v_1\})$
% and $\mathcal {B}(\{ v_1\})$ 
% the source and target vertices of the shortest paths
% cannot be $v_1$). 
% which is equal to $\mathcal \sigma(v_1) + \mathcal \sigma(v_2)- 2 \times \mathcal \sigma (\{ v_1,v_2\})$, an this is equal to
% $|\mathcal S(v_1) \xor \mathcal S(v_1)|$. 

\subsection{Computing exclusive betweenness centrality}
% 
% In this section, we discuss how exclusive betweenness can be computed.
A naive approach to compute exclusive betweenness centrality
of a given set is to enumerate all shortest paths
of the graph one-by-one, and check which one
passes over exactly one of the vertices in $A$.
% Then, those shortest paths that pass over more than one vertex in $A$ are detected and discarded.
Since the number of all shortest paths of the graph
(and the number of shortest paths 
that pass over exactly one of the vertices in $A$) is
exponential in the worst case (in terms of $n$),
this approach gives a worst case exponential time algorithm.

A more practical approach is to use the following inclusion-exclusion relationship:
\begin{align}
\mathcal{XB}(A) &= \sum_{v_i \in A} \mathcal{B}_A(v_i) - 2\sum_{\text{subsets } \{u,v\} \text{ of size 2 of }A} \mathcal{CC}_A(\{u,v\})\nonumber \\
      &+ 3\sum_{\text{subsets } \{u,v,w\} \text{ of size 3 of }A}\mathcal{CC}_A(\{u,v,w\}) - \hdots
\end{align}

In this approach, exclusive betweenness centrality of $A$ is 
computed based on betweenness centrality of the individual vertices in $A$ and co-betweenness centrality of the (larger) subsets of $A$.
While co-betweenness centrality of each subset of $A$ with an odd size contributes positively,
the contribution of each subset whose size is even is negative.
Using the method described in \cite{DBLP:conf/wsdm/Chehreghani14},
co-betweenness centrality of a set of vertices can be computed efficiently (in a low degree polynomial time, 
in terms of $m$ and $n$).
Overall, if the size of $A$ is considered as a constant,
since in this approach a constant number of
betweenness/co-betweenness scores will be computed, 
its time complexity will be polynomial in terms of $n$ and $m$
($O(nm)$ for unweighted graphs and
$O(nm+n^2\log n)$ for weighted graphs with positive weights).
Obviously, since the number of subsets of $A$ is exponential,
time complexity of this approach is exponential
in terms of $|A|$.

\subsection{Approximate algorithms}
\label{sec:approximate}

For large real-world networks consisting of thousands or millions of vertices, exact algorithms for computing centrality scores 
are usually intractable in practice.
Therefore, in recent years several approximate algorithms have been
developed for them.
An extensive study of approximate algorithms for
group betweenness centrality can be found in \cite{DBLP:conf/bigdataconf/ChehreghaniBA18}.
In the following, we investigate how these techniques
can be revised to compute
{\em exclusive betweenness centrality}.

\begin{algorithm}
\caption{High level pseudo code of the algorithm of estimating exclusive betweenness centrality.}
\label{algorithm:betweenness}
\begin{algorithmic} [1]
\STATE \textsc{ApproximateExclusiveBetweenness}
\STATE \textbf{Input.} A graph $G$, a non-empty set $A \subset V(G)$, and the number of samples $T$.
\STATE \textbf{Output.} Exclusive betweenness centrality of $A$.
\STATE $\beta \leftarrow 0$.
\STATE Compute probabilities $p_{ij}$, for each pair $(i,j) \in \mathcal N$.
\FORALL{$t=1$ \textbf{to} $T$ } \label{line:loop1}
\STATE Select a pair $(i,j) \in \mathcal N$ with probability $p_{ij}$.
% \STATE Form the SPD $D$ rooted at $i$
\STATE Let $\Pi_{ij}$ be the set of all shortest paths between $i$ and $j$.
Compute Probabilities $q_k$, for each $\pi_k \in \Pi_{ij}$.
\STATE Select a shortest path $\pi_k \in \Pi_{ij}$, with probability $q_k$.
\IF{exactly one of the members of $A$ are on $\pi_k$}
\STATE $\beta_t \leftarrow \frac{1}{p_{ij} \cdot q_k}$.
\STATE $\beta \leftarrow \beta + \frac{\beta_t}{T}$.
\ENDIF
% \STATE Compute dependency score of $i$ and $j$ on $A$, $\Delta_{ij}(A)$.
% \STATE $\beta_t \leftarrow \frac{\Delta_{ij}(A)}{p_{ij}}$.
% \STATE $\beta \leftarrow \beta + \frac{\beta_t}{T}$.
\ENDFOR \label{line:loop2}
\RETURN $\beta$.
\end{algorithmic}
\end{algorithm}

Algorithm~\ref{algorithm:betweenness} shows the high level pseudo code of a general algorithm for estimating exclusive
betweenness centrality.
It is similar to the general algorithm we presented in \cite{DBLP:conf/bigdataconf/ChehreghaniBA18}
for group betweenness centrality.
The key difference is that when a shortest path is sampled,
in Algorithm~\ref{algorithm:betweenness} it is checked whether \textbf{exactly one} of the vertices of $A$
are on the shortest path.
In estimating group betweenness centrality \cite{DBLP:conf/bigdataconf/ChehreghaniBA18},
it is checked whether \textbf{at least one} of the vertices of $A$
are on the shortest path.
Let $\mathcal N$ be the set of pairs in
$\left(V(G)\setminus A\right)\times\left(V(G)\setminus A \right)$.
% which forms one part of the sample space of our algorithm.
The input parameters of the algorithm are the graph $G$,
the set $A$ for which we want to estimate exclusive betweenness  centrality, and
the number of samples (iterations) $T$.
First, Algorithm~\ref{algorithm:betweenness} computes 
probabilities $p_{ij}$, for each pair $(i,j) \in \mathcal N$.
The probabilities $p_{ij}$ must satisfy the following conditions:
i) for each $(i,j) \in \mathcal N$, $p_{ij} > 0$, and
% such that
ii) $\sum_{(i,j) \in \mathcal N} p_{ij} = 1$.
Then, at each iteration $t$ of the loop in Lines \ref{line:loop1}-\ref{line:loop2} of
Algorithm \ref{algorithm:betweenness}:
\begin{itemize}
\item 
a pair $(i,j) \in \mathcal N$ is selected with probability $p_{ij}$,
\item
% the dependency score of vertices $i$ and $j$ on $S$, $\Delta_{ij}(S)$, is computed,
let $\Pi_{ij}$ be the set of all shortest paths between $i$ and $j$.
Probabilities $q_k$ are computed, for each shortest path $\pi_k \in \Pi_{ij}$, 
\item
a shortest path $\pi_k$ from $i$ to $j$ is selected with probability $q_k$,
\item
if \textbf{exactly one} of the members of $A$ are on $\pi_k$,
$\beta_t$,
the estimation of $\mathcal{XB}(A)$ at iteration $t$,
is defined as
$$\frac{1}{ p_{ij} \cdot q_k}.$$
Otherwise, it is defined as $0$.
% $D$ is traversed in a bottom-up way and
% betweenness centrality of every vertex is estimated as
% $ \sum_{w:u \in P_i(w)} \frac{\sigma_{su}}{\sigma_{sw}}\left( \frac{1}{p_i} + BC(w) \right) $.
\end{itemize}

The average of exclusive betweenness scores estimated at different iterations is returned as
the final estimation of the exclusive betweenness centrality of $A$.
In a way similar to Lemma~1 of \cite{DBLP:conf/bigdataconf/ChehreghaniBA18},
it can be shown that $\beta$ yields an unbiased estimation of
exclusive betweenness centrality of $A$, i.e.,
the expected value of $\beta$ is equal to the
exclusive betweenness centrality of $A$.

Note that while Algorithm~\ref{algorithm:betweenness} estimates exclusive betweenness centrality of a given set $A$,
it can be simply revised to
estimates exclusive betweenness centrality of
all (non-empty and proper) subsets of the vertices of $G$.
To do so, at each iteration $t$ and after sampling
the shortest path $\pi_k$,
the exclusive betweenness score of any (non-empty) subset of
the vertices of $G$ that satisfies both of the following conditions:
\begin{itemize}
\item
its \textbf{exactly one member}
is an internal vertex of $\pi_k$, and
\item
none of its members are either the source or the target of $\pi_k$, 
\end{itemize}
% internal vertices of $\pi_k$
is estimated as $$\frac{1}{ p_{ij} \cdot q_k}$$
and any other subset of $V(G)$ as $0$.
The final estimation of the
exclusive betweenness centrality of each subset is 
the average of its estimated exclusive betweenness scores,
at different iterations.
% In this case, Lemma~\ref{lemma:expectedvalue} will be valid
% for any (non-empty and proper) subset of the vertices of $G$. 
The key difference between this case and the case of
estimating group betweenness centrality of all subsets of the vertices
is as follows:
for group betweenness centrality, as mentioned in \cite{DBLP:conf/bigdataconf/ChehreghaniBA18},
at each iteration $t$ and after sampling the shortest path $\pi_k$,
group betweenness score of any (non-empty) subset of 
the vertices of $G$ 
whose \textbf{at least one member}
is an internal vertex of $\pi_k$,
is estimated as $$\frac{1}{ p_{ij} \cdot q_k},$$
and any other subset of $V(G)$ as $0$. \footnote{Note that since
in \cite{DBLP:conf/bigdataconf/ChehreghaniBA18}
group betweenness centrality of set $A$ is defined as 
$$\sum_{s,t \notin A: s\neq t}
\frac{\sigma^*_{st}(A)}{\sigma_{st}},$$
its estimation 
at iteration $t$
is: 
$$\frac{1}{|\Pi_{ij}| \cdot p_{ij} \cdot q_k}.$$}

In the following,
we present some specific forms of the above mentioned general algorithm.
\begin{itemize}
\item
In a {\em source vertex sampling} algorithm,
at each iteration,
first a source vertex $i$ is sampled with probability $p_i$.
Then, the number of shortest paths that start from $i$
and pass over exactly one of the members of $A$, is counted.
Then, this number is divided by $p_i$
to yield the estimation of the exclusive betweenness centrality of $A$
at the current iteration.
In the end, the final estimation is the average of estimations
of different iterations.
In a special case of this algorithm,
called {\em uniform source vertex sampling} algorithm,
for each vertex $i\in V(G)\setminus A$,
$p_i$ is defined as $$\frac{1}{|V(G)\setminus A|};$$
and for each vertex $i \in A$,
it is defined as $0$.
\item
In a {\em pair sampling} algorithm,
at each iteration,
first a source vertex $i$ and a target vertex $j$ 
(a pair of vertices $i,j$)
are sampled with probability $p_{ij}$.
Then, the number of shortest paths that start from $i$,
pass over exactly one of the members of $A$
and end to $j$ is counted.
Then, this number is divided by $p_{ij}$
to yield the estimation of the exclusive betweenness centrality of $A$
at the current iteration.
In the end, the final estimation is the average of estimations
of different iterations.
In a special form of this algorithm,
called {\em uniform pair sampling} algorithm,
for each pair of vertices $i,j \in V(G)\setminus A \times \in V(G)\setminus A$ ($i \neq j$), 
$p_{ij}$ is defined as
$$\frac{1}{(|V(G)\setminus A|)(|V(G)\setminus A|-1)}$$
and for any other pair,
it is defined as $0$.
\item
In a form of {\em shortest path sampling} algorithm,
at each iteration,
first a pair of vertices $i,j \in (V(G)\setminus A) \times (V(G)\setminus A)$ so that $i \neq j$ 
are sampled uniformly at random.
Then, one of the shortest paths from $i$ to $j$
is sampled uniformly at random.
Then, it is checked whether exactly one of the 
vertices in $A$ is an internal vertex of the sampled path. 
The number such {\em occurrences} during different iterations is counted.
In the end, this count is scaled to 
give an unbiased estimation of the 
exclusive betweenness centrality of $A$.
% vertices in $A$ are an internal vertexof the sampled path. 
% the number of shortest paths that start from $i$,
% pass over exactly one of the members of $A$
% and end to $j$ is counted.
% Then, this number is devided by $p_{ij}$
% to yield the estimation of the exclusive betweenness centrality of $A$
% at the current iteration.
% In the end, the final estimationis the average of estimations
% of different iterations.
% In a special case of this algorithm,
% called {\em uniform source vertex sampling} algorithm,
% for each vertex $i\in V(G)\setminus A$,
% $p_i$ is defined as $\frac{1}{|V(G)\setminus A|}$;
% and for each vertex $i \in A$,
% it is defined as $0$.
\end{itemize}

\section{Experimental results}
\label{sec:experimentalresults}

In this section, we empirically analyze exclusive betweenness centrality.
First, we evaluate running time of computing exclusive betweenness centrality
over a number of real-world networks.
Then, we investigate the correlation between exclusive betweenness centrality
and the other centrality notions of sets,
such as group betweenness centrality and co-betweenness centrality.
% We performed  experiments on real-world networks from different domains to assess the quantitative and qualitative behavior of
% the proposed algorithm.
The experiments are done on one core of a single AMD
Processor
% 270 clocked at 2.0 GHz
with 4 GB
main memory.
% and $2\times1$ MB L2 cache, running Ubuntu
% Linux 12.0.
% The program was compiled by the GNU C++ compiler 4.0.2 using optimization level 3.

\subsection{Empirical evaluation of exclusive betweenness centrality computation}

We evaluate the empirical efficiency of the exact algorithm of
computing exclusive betweenness centrality, discussed in Section~\ref{sec:algorith}. 
% computation over several real-worldnetworks.
% performed  experiments on real-world networks from different domains to assess the quantitative and qualitative behavior of
% the proposed algorithm.
We test the algorithm over six real-world datasets.
Table \ref{table:datase} summarizes specifications of the datasets.
Figure~\ref{fig:time} presents the running times (for different set sizes).
Over each dataset, we considered different set sizes varying from 
$2$ to $5$.
For each set size $k$, we select $50$ random subsets of the
vertices of size $k$,
and compute their exact exclusive betweenness scores.
In the end, for each set size, we report in Figure~\ref{fig:time}
the running time of 
the set that takes the longest time.
As can be seen in the figure, by increasing the set size,
running time gradually increases.
The reason is that as discussed in Section~\ref{sec:algorith},
by increasing the set size more co-betweenness centralities are required 
to be computed.
This increases the run time of computing exclusive betweenness centrality.
This is unlike the run time of computing co-betweenness centrality
which as discussed in \cite{DBLP:conf/wsdm/Chehreghani14},
usually decreases by increasing the size of the set.

\begin{table*}
\caption{\label{table:datase}Summary of real-world networks.}
\begin{center}
\begin{tabular}{ l l l l l}
\hline
Dataset & \# vertices & \# edges & maximum degree &  URL \\ \hline
jazz\cite{nr-aaai15}  & 198 & 2.7K & 100 &\url{http://networkrepository.com/jazz.php}  \\ 
bwm200 \cite{nr-aaai15} & 200 & 596 & 6 & \url{http://networkrepository.com/bwm200.php} \\ 
can\_187 \cite{nr-aaai15}  & 187 & 652  & 9& \url{http://networkrepository.com/can-187.php}  \\
can\_256 \cite{nr-aaai15}  & 256 & 1.3K & 82 & \url{http://networkrepository.com/can-256.php}  \\
ca-netscience \cite{nr-aaai15,newman2006finding}  & 379 & 914 & 34 &\url{http://networkrepository.com/ca-netscience.php}  \\
GD00\_c \cite{nr-aaai15}  & 638 & 1K & 58 &\url{http://networkrepository.com/GD00-c.php}  \\
\hline
\end{tabular}
\end{center}
\end{table*}

\begin{figure*}
\centering
\subfigure[bwm200]
{
\includegraphics[scale=0.35]{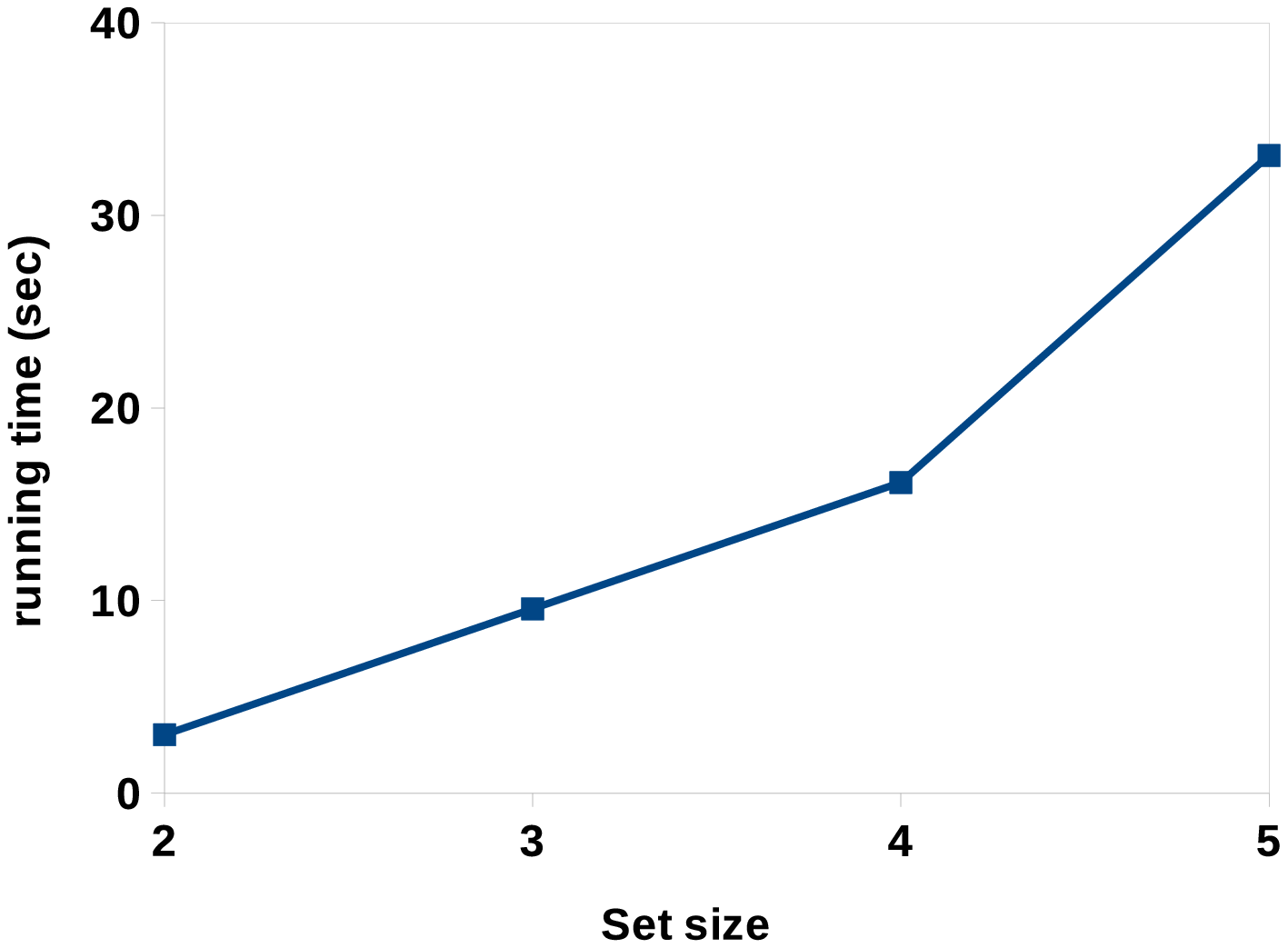}
\label{fig:alpha}
}
\subfigure[can\_187]
{
\includegraphics[scale=0.35]{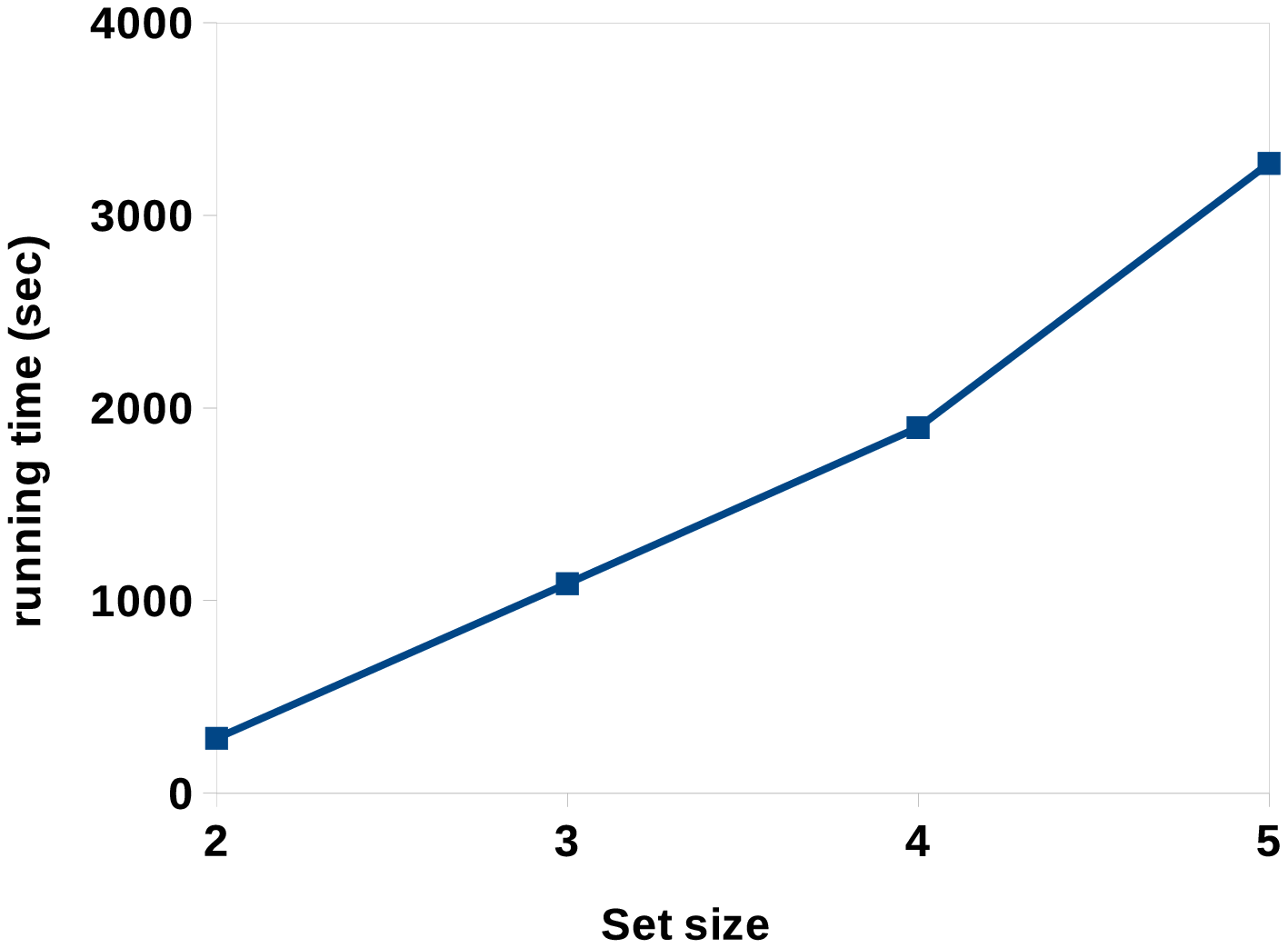}
\label{fig:time}
}
\subfigure[can\_256]
{
\includegraphics[scale=0.35]{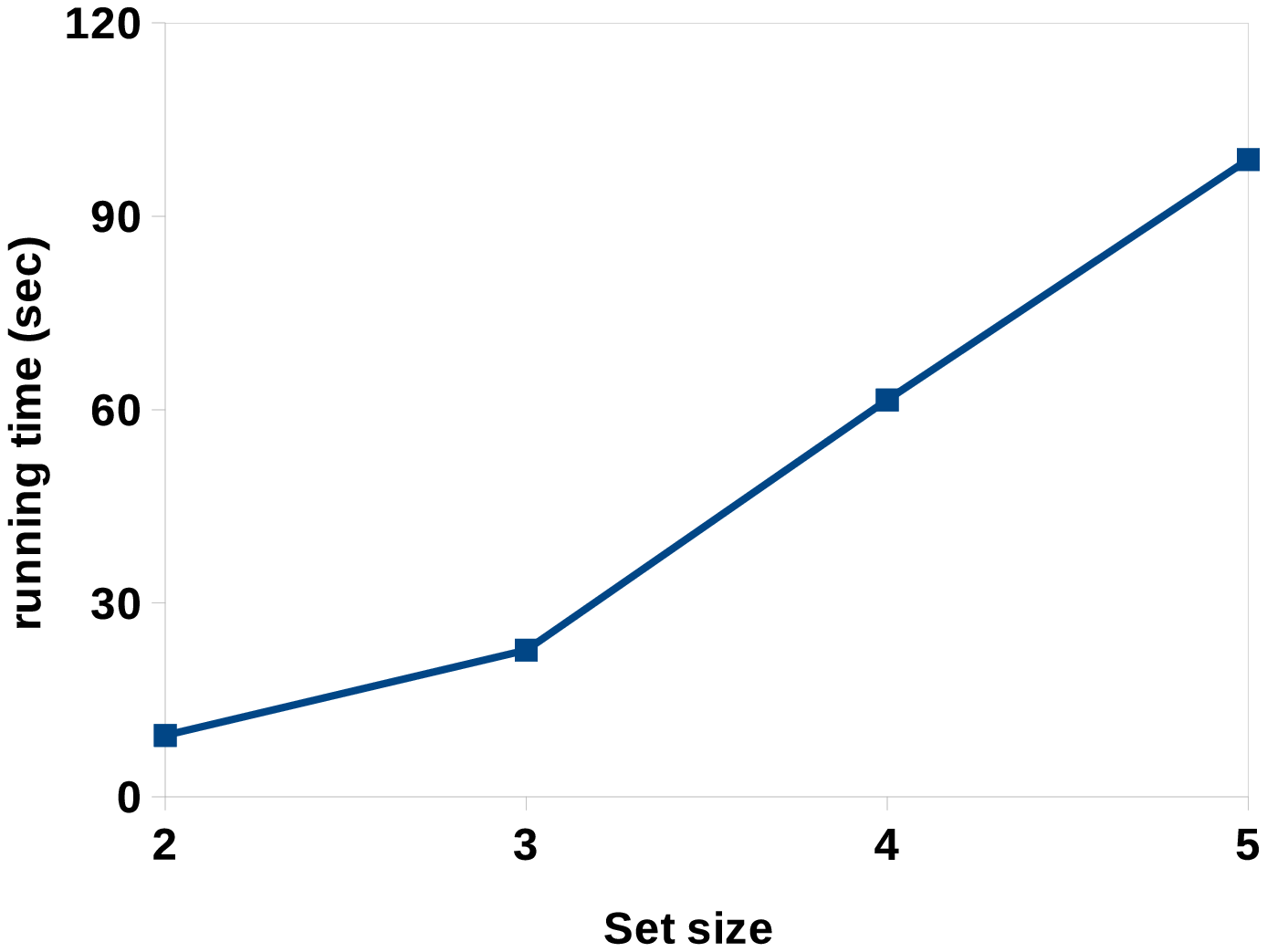}
\label{fig:time}
}
\subfigure[ca-netscience]
{
\includegraphics[scale=0.35]{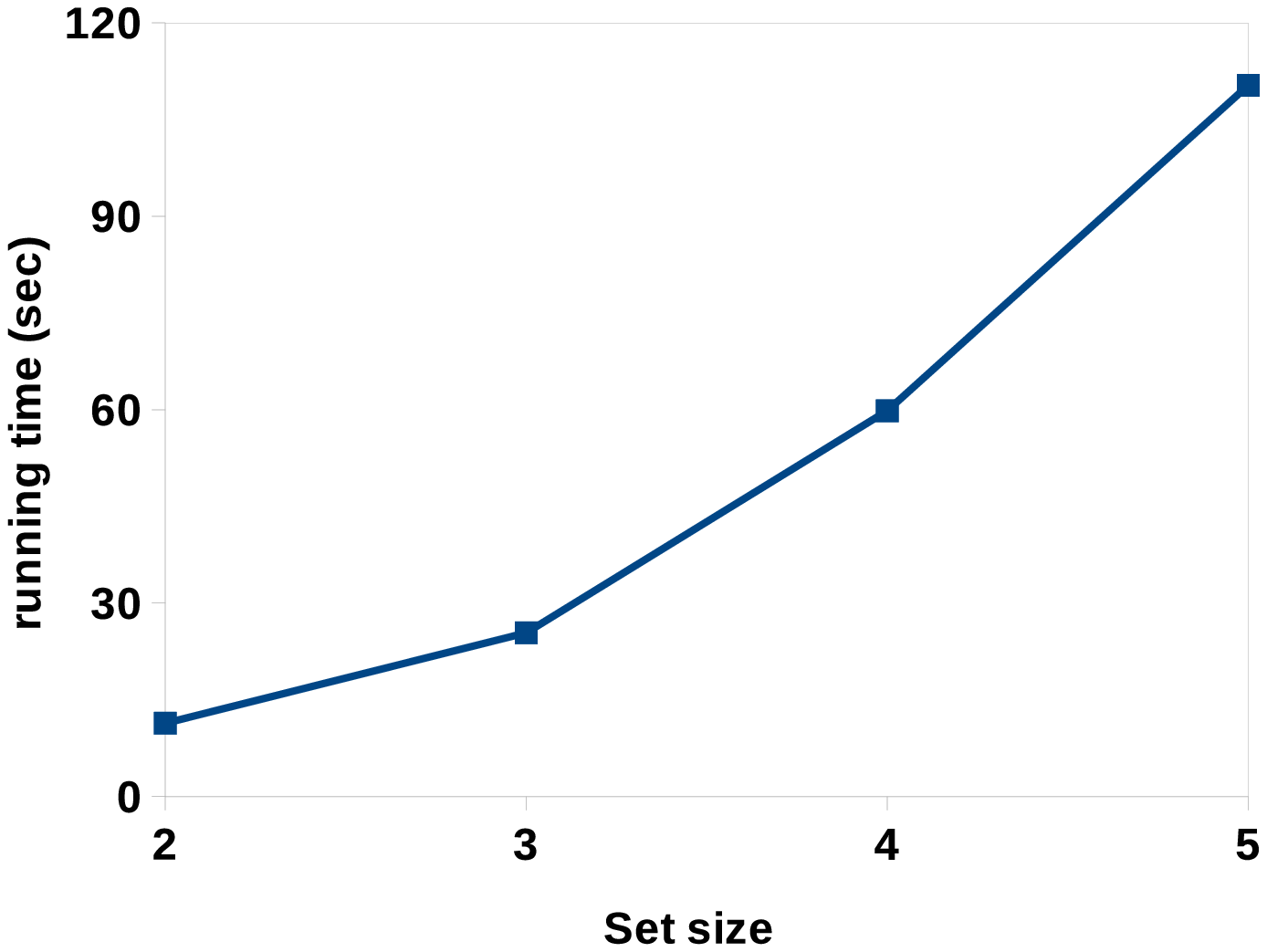}
\label{fig:time}
}
\subfigure[GD00\_c]
{
\includegraphics[scale=0.35]{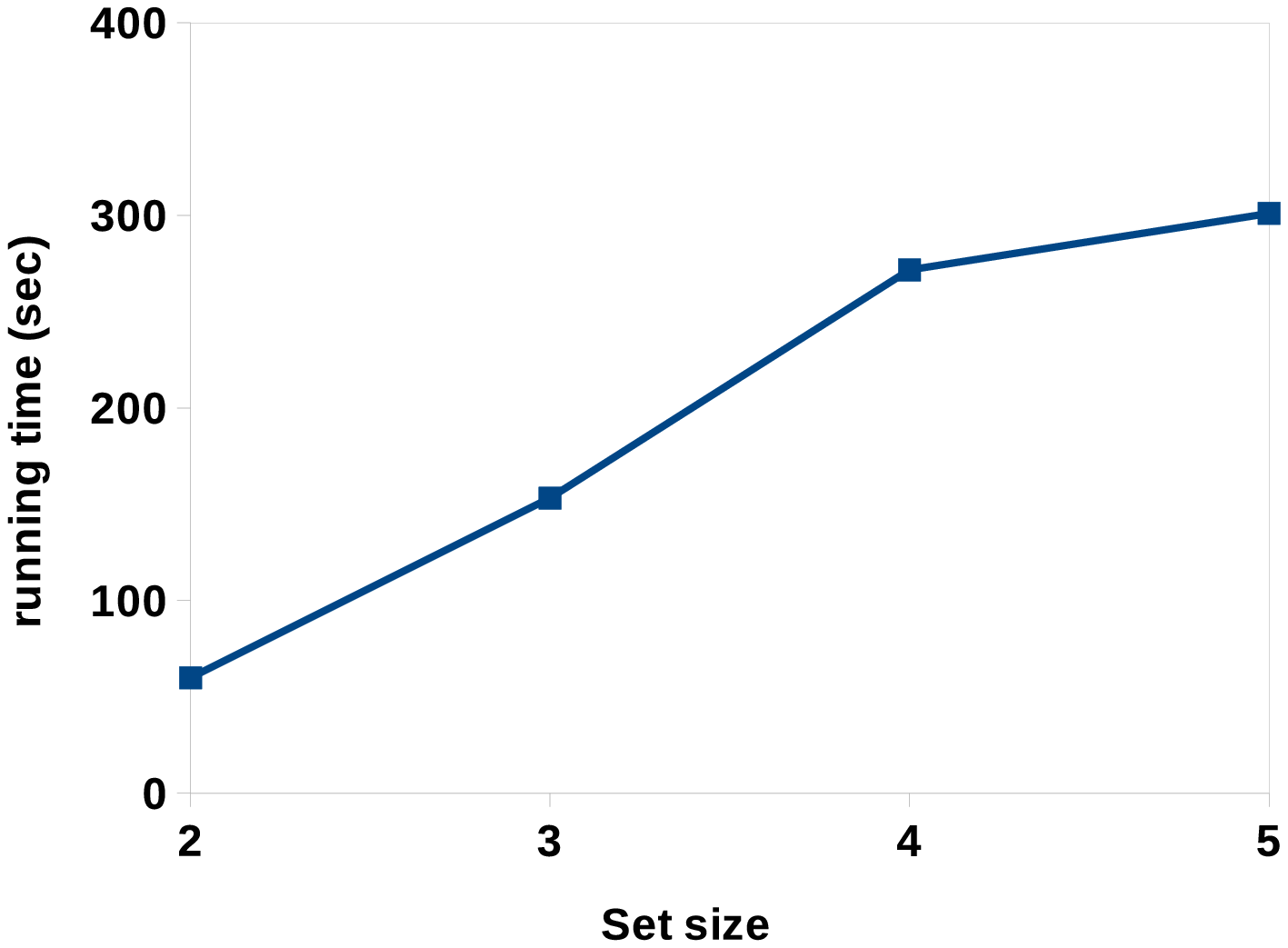}
\label{fig:time}
}
\subfigure[jazz]
{
\includegraphics[scale=0.35]{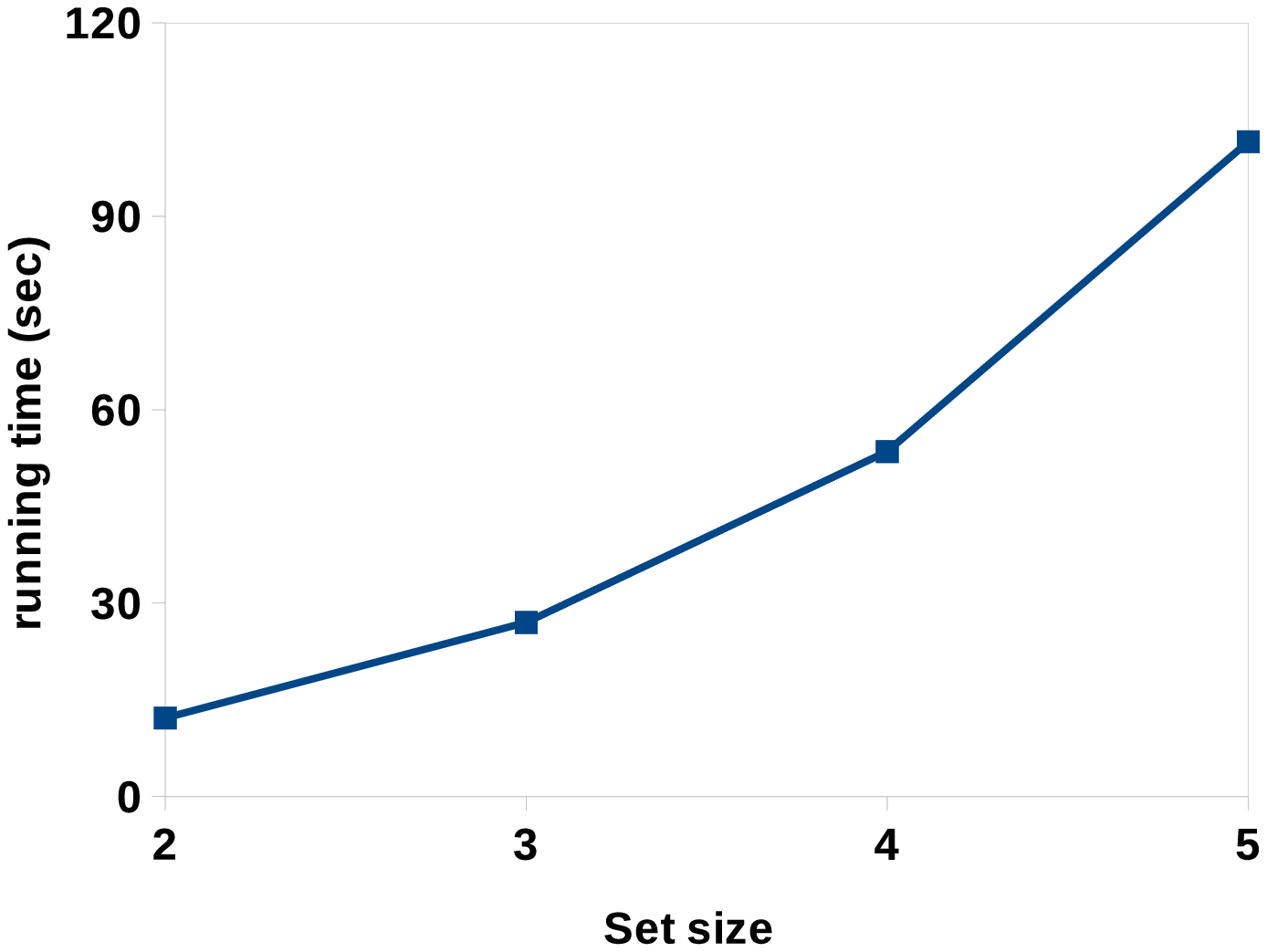}
\label{fig:time}
}
\caption
{
\label{fig:time}
Running time of the exclusive betweenness centrality 
computation algorithm over different datasets.}
\end{figure*}

\subsection{Correlation with other set centrality notions}

\begin{figure}
\centering
\includegraphics[scale=0.17]{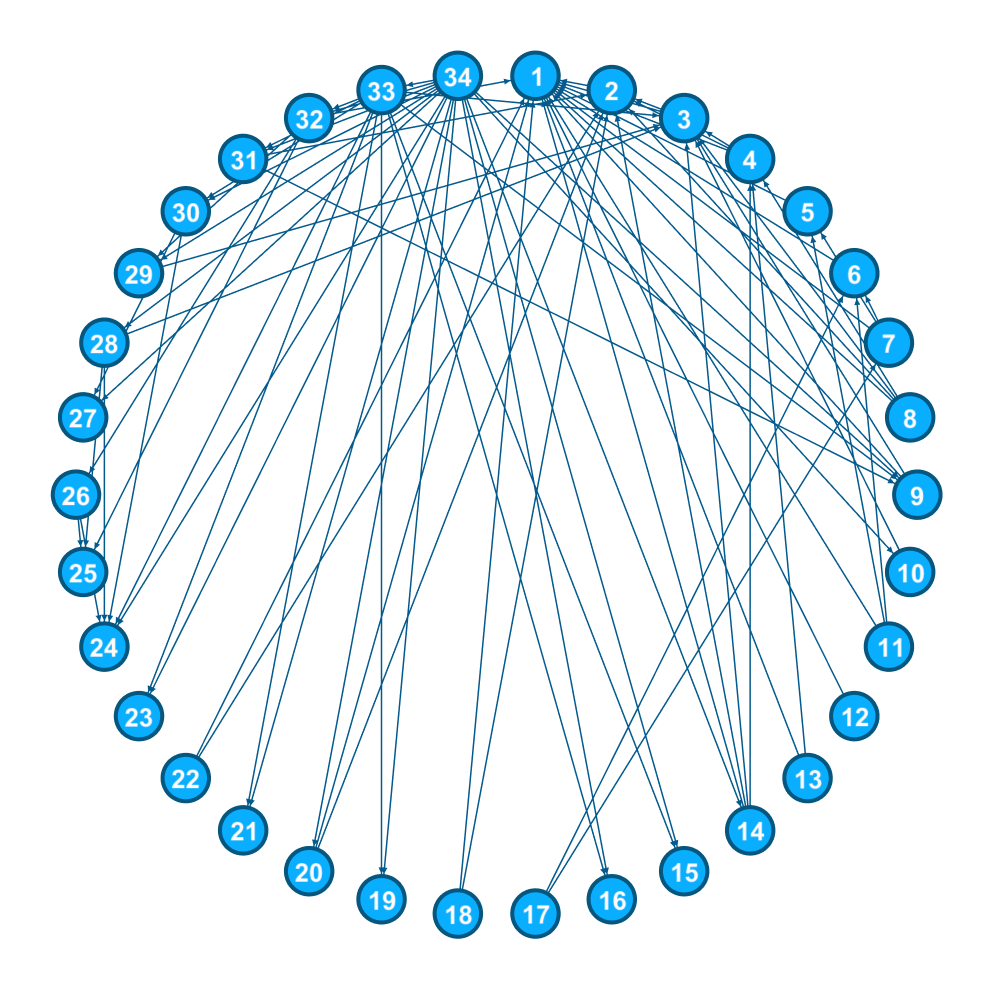}
\caption
{
\label{fig:zachary}
The Zachary karate club network.
Source: Wikipedia
% \url{https://en.wikipedia.org/wiki/Zachary%27s_karate_club}
}
\end{figure}

In this section, we investigate the correlation between
exclusive betweenness centrality and
group betweenness centrality and co-betweenness centrality.
We examine the centrality notions on the well-known
Zachary karate club network \cite{Zac77}.
Zachary collected this dataset from the members of a university karate club.
In this undirected network, each vertex represents a member of the club,
and each edge represents a relationship between two members of the club.
% An often discussed problem using this dataset is to find the two groups of people into which the karate club split after an argument between two teachers.
It has $34$ vertices, $78$ edges, its maximum degree $17$,
and its diameter is and $5$.
This network is depicted in Figure~\ref{fig:zachary}.

\begin{figure*}
\centering
\subfigure[The correlation between exclusive and group betweenness centralities.]
{
\includegraphics[scale=0.45]{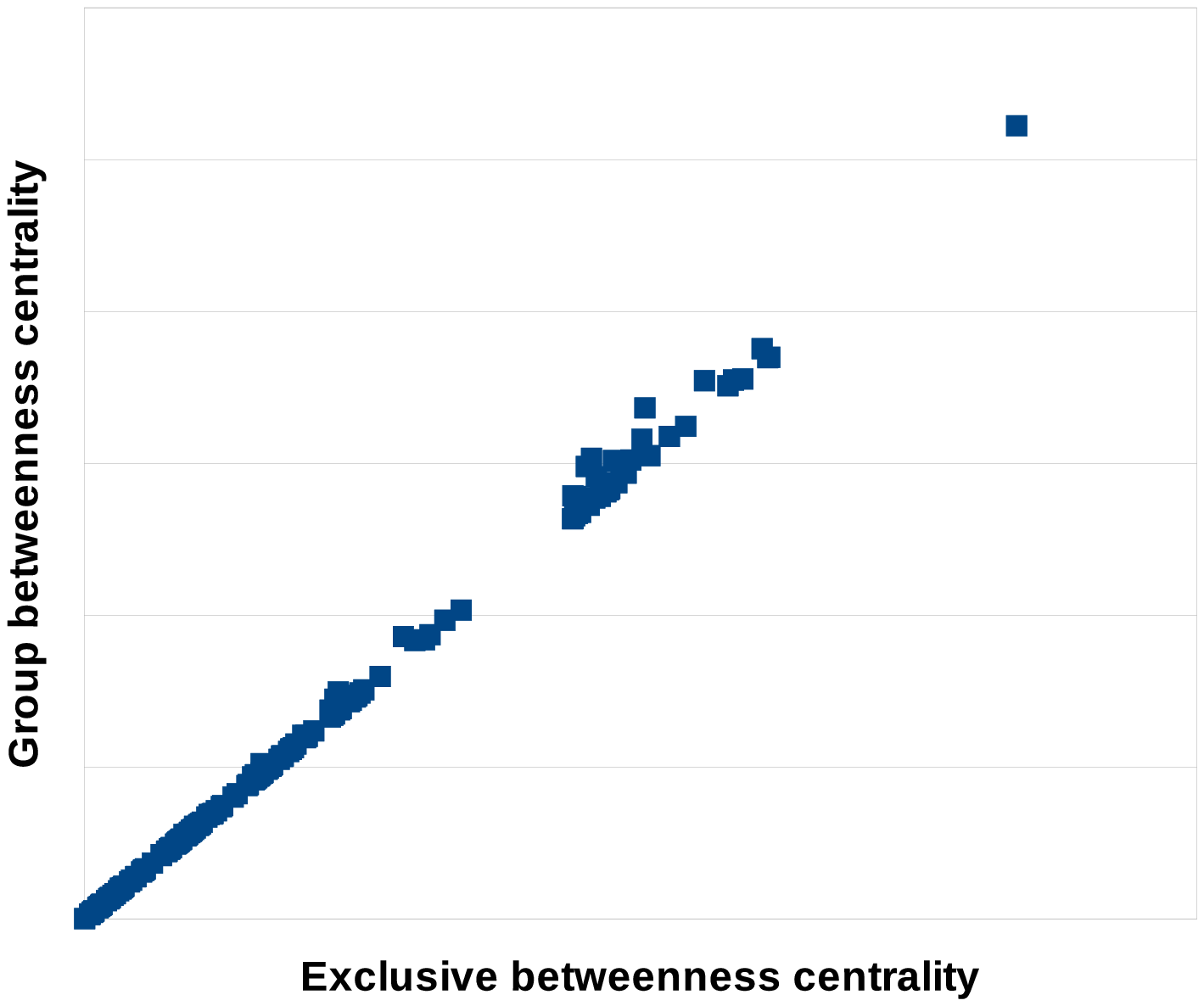}
\label{fig:xb-gb}
}
\subfigure[The correlation between exclusive and co-betweenness centralities.]
{
\includegraphics[scale=0.45]{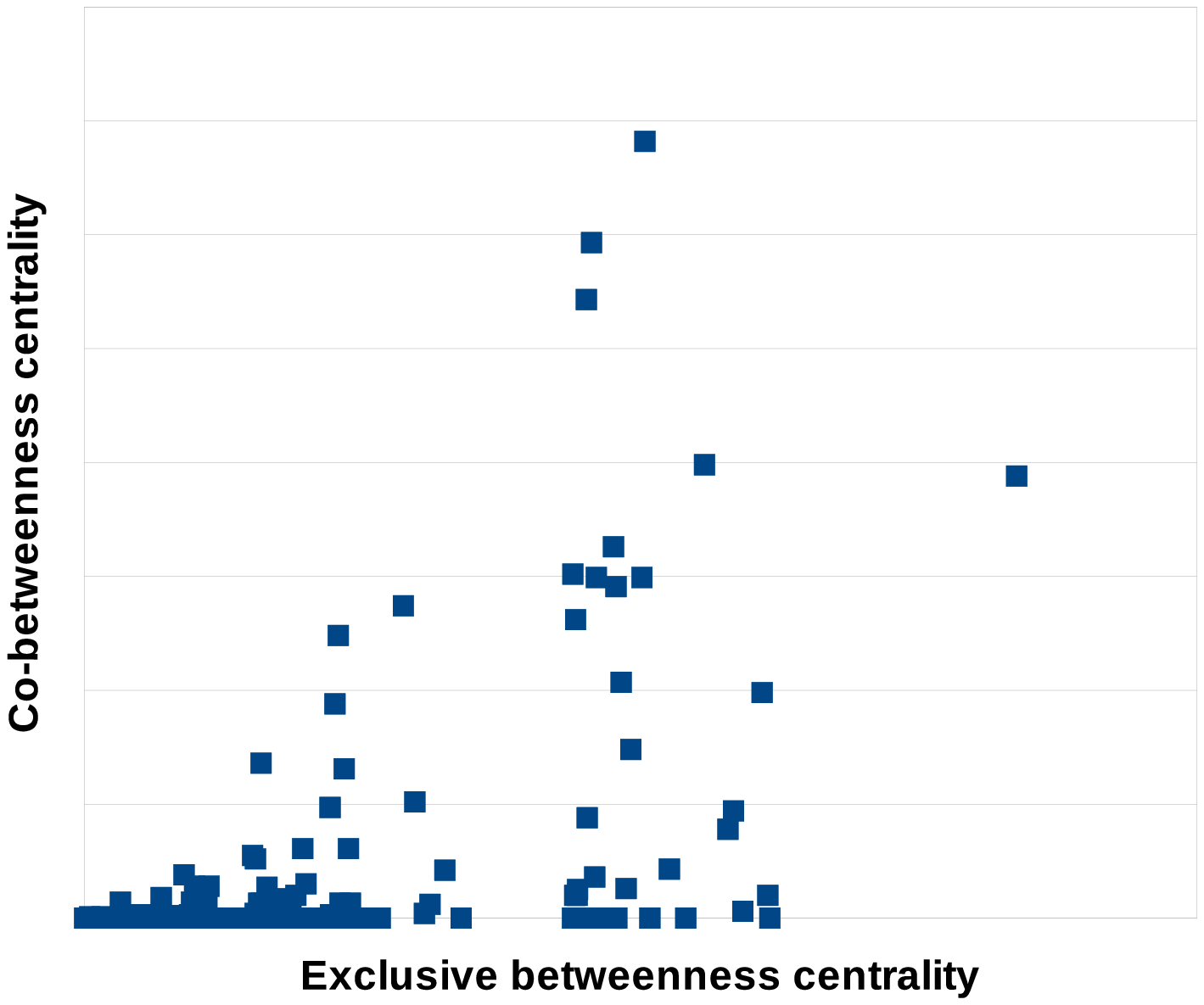}
\label{fig:xb-cb}
}
\caption
{
\label{fig:correlation}
Correlations between exclusive betweenness centrality and the other set betweenness centralities.
}
\end{figure*}

Figure~\ref{fig:correlation}
represents the correlations, wherein the centrality values of
all the sets of size $2$ are examined.
As can be seen in the figure,
there is almost a linear correlation between
exclusive betweenness centrality and group betweenness centrality,
so that sets with large group betweenness centrality
have also a large exclusive betweenness centrality,
and vice versa.
However, the correlation between exclusive betweenness centrality and 
co-betweenness centrality is not direct and regular,
as having a high exclusive betweenness score does not always imply 
a high co-betweenness centrality.

\section{Related Work}
\label{section:relatedwork}

Centrality measures are important and essential tools for analyzing social and information networks. 
some of widely used indices for centrality are \textit{betweenness centrality} \cite{jrnl:Freeman}, \textit{closeness centrality} \cite{DBLP:journals/cj/ChehreghaniBA18}, \textit{degree centrality} \cite{book:Wasserman}, \textit{eigenvector centrality} \cite{jrnl:Bonacich} and PageRank \cite{Arasu:2002:PCS}.
Betweenness centrality, which is widely used as a precise estimation of the information flow controlled by a vertex in social and information networks, assumes that information flow is done through shortest paths \cite{jrnl:Yan}.
Barthelemy \cite{Barthelemy:643484} showed that
many {\em scale-free networks} \cite{jrnl:Barabasi,DBLP:conf/uai/ChehreghaniC16,DBLP:journals/ipl/ChehreghaniA17} have a power-law distribution of betweenness Centrality.
Brandes~\cite{jrnl:Brandes}  introduced a new algorithm for computing betweenness centrality of a vertex,
which is performed in $O(nm)$ time and $O(nm + n^2 \log n)$ time for unweighted networks and (positively) weighted networks, respectively.
In recent years several exact and approximate algorithms
have been proposed to improve the efficiency of betweenness centrality computation \cite{DBLP:journals/cj/Chehreghani14,DBLP:conf/pakdd/ChehreghaniBA18,DBLP:conf/bigdataconf/ChehreghaniBA18a,DBLP:conf/edbt/ChehreghaniAB19,DBLP:conf/cikm/ChehreghaniBA19}. 

Everett and Borgatti \cite{citeulike:392816} defined group betweenness centrality as a natural extension of betweenness centrality for sets of vertices.
% Group betweenness centrality of a set is defined as the number of shortest paths passing through at least one of the vertices in the set \cite{citeulike:392816}.
The authors of \cite{DBLP:conf/bigdataconf/ChehreghaniBA18} provided an extensive comparison of different group betweenness centrality estimation algorithm.
They also presented an extension of distance-based sampling
for group betweenness centrality.
The other natural extension of betweenness centrality is \textit{co-betweenness centrality}.
Co-betweenness centrality is defined as the number of shortest paths passing through all vertices in the set \cite{jrnl:Kolaczyk}.
The authors of \cite{jrnl:Kolaczyk} proposed an algorithm for individual co-betweenness centrality computation,
that works only for sets of size $2$ and its time complexity is $O(n^3)$.
Chehreghani \cite{DBLP:conf/wsdm/Chehreghani14} presented 
algorithms for co-betweenness centrality computation of a
set of an arbitrary size, and showed that by increasing the size of the set, its co-betweenness centrality can be computed 
more efficiently.
Time complexity of these algorithms is $O(nm+n^2\log n)$ or less.

% Puzis et al. \cite{jrnl:PuzisPhysRev} proposed a $O(|K|^3)$ time algorithm for computation of successive group betweenness centrality.
% In \cite{jrnl:PuzisAIComm}, the authors presented two algorithms for
% finding the most prominent group.
% The first algorithm is based on heuristic search and the second is based on iterative greedy choice of vertices.
The authors of \cite{DBLP:journals/jacm/DolevEP10} 
presented the Routing Betweenness Centrality (RBC) index
and proposed algorithms for computing RBC of individual vertices
and algorithms
for computing group RBC of a given set (or sequence) of vertices. 
Ballester et.al. \cite{article_Econometrica} discussed the importance of finding the key group in a criminal network.
Borgatti \cite{10.1007/s10588-006-7084-x} discussed that the Key Player Problem (KPP) is strongly related to the cohesion of a network.
He introduced two problems: KPP-Pos and KPP-Neg.
He showed that the solution of KPP-Pos is a group maximally connected to all other vertices in a graph and the solution of KPP-Neg 
is a group maximally disrupting the network.

\section{Conclusion}
\label{section:conclusion}

In this paper, we suggested generalizing
the function of aggregating the centralities of the vertices of a set,
to obtain the centrality of the whole set.
As a particular case, we studied 
exclusive (betweenness) centrality,
wherein the number of shortest paths that pass
over exactly one of the vertices in the set,
is counted.
We also presented exact and approximate
algorithms for computing 
exclusive betweenness centrality, efficiently.
By conducting extensive experiments,
% In our experiments, first we
first we evaluated the empirical efficiency
of exclusive betweenness centrality computation.
Then, we
investigated
the correlations between exclusive betweenness centrality
and the other set centrality notions.

\bibliographystyle{plain}
\bibliography{allpapers}   % name your BibTeX data base

% \bibliographystyle{named}
% \end{thebibliography}

\end{document}